\documentclass[10pt,aps,prd,twocolumn,showpacs,amsmath,amssymb,nofootinbib,eqsecnum,preprintnumbers,longbibliography]{revtex4-1}

\usepackage[english]{babel}				
\usepackage[mathscr]{eucal} 			
\usepackage{accents}					
\usepackage{mathtools}					
\usepackage[ddmmyyyy]{datetime}    		
\usepackage{enumitem}					
\usepackage{xcolor}						

\newcommand\bs[1]{\boldsymbol{#1}}

\newcommand\dd{\mathrm{d}}
\newcommand\pp{\partial}
\newcommand\de{\delta}

\newcommand\feq{\mathrel{\phantom{=}}}

\newcommand{\bvee}{\bs{\vee}}

\newcommand{\bdot}{\bs{\cdot}}

\DeclareMathOperator\erf{erf}

\begin{document}


\title{NUT charge in linearized infinite derivative gravity}

\author{Ivan Kol\'a\v{r}}
\email{i.kolar@rug.nl}
\affiliation{Van Swinderen Institute, University of Groningen, 9747 AG, Groningen, The Netherlands}

\author{Anupam Mazumdar}
\email{anupam.mazumdar@rug.nl}
\affiliation{Van Swinderen Institute, University of Groningen, 9747 AG, Groningen, The Netherlands}

\date{\today}

\begin{abstract}
We study the gravitational field of the NUT-like source in the linearized (ghost-free) infinite derivative gravity. Such a source is equivalent to the spinning semi-infinite cosmic string with no tension. In general relativity, the linearized (massless) Taub--NUT solution has a curvature singularity as well as a topological defect corresponding to distributional curvature on one half of the symmetry axis called the Misner string. We find the NUT-charged spacetime in the linearized infinite derivative gravity. We show that it is free from curvature singularities as well as Misner strings. We also discuss an asymptotic limit along the symmetry axis that leads to the spacetime of a spinning cosmic string of infinite length.
\end{abstract}

\maketitle


\section{Introduction}

General relativity is a very successful theory of gravity at the scales of our solar system \cite{Will:2014kxa}. However, the theory is incomplete in the ultraviolet regime, i.e., for very short distances and time intervals. It contains black-hole and cosmological singularities and fails to be perturbatively renormalizable at the quantum level. It is well known that if quadratic terms in the curvature are added to the Einstein--Hilbert action \cite{Stelle:1977ry}, the resulting gravitational theory is renormalizable \cite{Stelle:1976gc}. Inclusion of further higher-order and higher-derivative terms leads to super-renormalizable models of quantum gravity \cite{Asorey:1996hz}. In addition, the gravitational potential becomes regular \cite{Modesto:2014eta,Giacchini:2018wlf}. Unfortunately, these theories suffer from the presence of ghost degree of freedom in the physical spectrum. 

The \textit{(ghost-free) infinite derivative gravity} theories provide an interesting solution to this problem. Their action involves non-local terms containing \textit{form-factors} with infinitely many derivatives (similar to those which appear frequently in effective descriptions of the string field theory \cite{Witten:1985cc} or the $p$-adic string theory \cite{Freund:1987kt,Freund:1987ck,Brekke:1988dg,Frampton:1988kr}). A proper choice of the form-factors ensures that no additional degrees of freedom appear in these theories (see, e.g., \cite{Biswas:2011ar,Biswas:2013kla,Biswas:2013cha,Modesto:2011kw,Modesto:2012ys}, or \cite{Tomboulis:1997gg}, for a model proposed earlier). Furthermore, the quantum aspects and the renormalizability of the infinite derivative gravity theories was discussed in \cite{Tomboulis:1997gg,Modesto:2014lga,Talaganis:2014ida,Tomboulis:2015esa,Modesto:2017hzl,Buoninfante:2018mre}. Considering that the quantum fluctuations appear at much larger scales than the scale of non-locality of these theories, the metric can be treated as classical. The studies from this classical point of view show that the infinite derivative gravity may actually resolve the cosmological, black-hole, and other gravitational singularities \cite{Biswas:2005qr,Koshelev:2017tvv,SravanKumar:2018dlo,Biswas:2010zk,Biswas:2012bp,Koshelev:2018rau,Biswas:2011ar,Frolov:2015usa,Frolov:2015bia,Frolov:2015bta,Edholm:2016hbt,Buoninfante:2018xiw,Koshelev:2018hpt,Buoninfante:2018rlq,Buoninfante:2018stt,Buoninfante:2018xif,Boos:2018bxf,Boos:2020kgj,Boos:2020ccj}.

In particular, it was shown that the non-locality admits a bouncing cosmological universe \cite{Biswas:2005qr}, inflationary solution \cite{Koshelev:2017tvv,SravanKumar:2018dlo}, while the Kasner-type solution with the anisotropic collapse is not permitted \cite{Koshelev:2018rau}. The bouncing solution turns out to be free from perturbative instabilities \cite{Biswas:2010zk,Biswas:2012bp}. In the context of black-hole singularities, it was argued that the Schwarzschild-type metric cannot be a vacuum solution of the infinite derivative gravity \cite{Buoninfante:2018rlq,Koshelev:2018hpt}. It is well known that the ${1/r}$ behavior of the gravitational potential of the point-like source is regularized in the linear theory \cite{Biswas:2011ar,Edholm:2016hbt} because the delta source is effectively smeared by the non-local operator. As a result, the curvature of the metric is finite everywhere \cite{Buoninfante:2018xiw}. A similar feature remains true for electrically charged sources \cite{Buoninfante:2018stt}, rotating ring-type sources \cite{Buoninfante:2018xif}, and other extended objects associated with topological defects such as the p-branes, cosmic strings, and gyratons \cite{Boos:2018bxf,Boos:2020kgj,Boos:2020ccj}. It was also demonstrated that there exists a mass gap for mini-black-hole production by a spherical gravitational collapse \cite{Frolov:2015bia,Frolov:2015bta} and head-on collision of ultrarelativistic particles \cite{Frolov:2015usa}, so the theory never develops a singularity at the linear level. Let us also note that the results in the linearized theory are actually more significant in the infinite derivative gravity than in the general relativity because the non-localities tend to weaken the gravitational interaction at short distances.

The Taub--NUT spacetime \cite{Taub:1950ez,Newman:1963yy} is arguably one of the most puzzling solutions of general relativity. It carries the \textit{NUT charge}, which is a gravitational analog to the magnetic monopole \cite{Dowker:1967zz}. The metric is endowed with a very peculiar type of singularity on the symmetry axis (similar to Dirac's string \cite{Dirac:1931kp}), called the \textit{Misner string}, that is surrounded by a region with closed timelike curves. There exist two prominent proposals for the interpretation of this geometry. In Misner's interpretation \cite{Misner:1963fr}, the string is rendered unobservable by assuming the periodicity in time. This approach not only leads to the existence of closed timelike curves in the whole spacetime, but it also causes severe issues with an analytic extension of the spacetime. In an alternative approach suggested by Bonnor \cite{Bonnor:1969} (see also \cite{Sackfield:1971,Bonnor:1992,Griffiths:2009dfa,Kolar:2019gzy}), the periodicity in time is abandoned, and the Misner string is treated as a topological defect caused by a linear material source of angular momentum. In recent years, the Taub--NUT spacetime with Bonnor's interpretation has received increasing interest because the significant obstructions to consider this solution unphysical have been removed. Specifically, it was shown that the Misner string is fully transparent for geodesics (which makes the spacetime geodesically complete) and there is no violation of causality for timelike and null geodesics \cite{Clement:2015cxa}.\footnote{A different approach to solving the problems with the Misner string was proposed in \cite{Gera:2019ebe}.} Another evidence supporting its possible physical significance is the recent construction of the consistent black-hole thermodynamics with NUT charge \cite{Kubiznak:2019yiu,Bordo:2019tyh}. The Taub--NUT metric was also studied in the context of the Kerr--Schild double copy \cite{Luna:2015paa}, where it was mapped to a dyon whose electric and magnetic charges copy to mass and NUT charge.

The aim of this paper is to extend the class of linearized solutions \cite{Biswas:2011ar,Buoninfante:2018stt,Buoninfante:2018xif,Boos:2018bxf,Boos:2020kgj} by finding an analytic solution describing the gravitation field of the NUT charge. Following the Bonnor's interpretation, we view the (massless) Taub--NUT solution in the linearized regime for small NUT charge as the spacetime of a spinning semi-infinite cosmic string. We show that the distributional source is smeared by the non-locality. The resulting solution is regular everywhere. In the local limit and far from the source, we recover the solution of general relativity. Exploring the asymptotic limit of the metric along the symmetry axis, we obtain the non-local solution for spinning cosmic string of infinite length.

The paper is organized as follows: In Sec.~\ref{sc:pre} we review the ghost-free infinite derivative gravity and some properties of the linearized Taub--NUT solution. The main results of the paper are in Sec.~\ref{sc:nut}, where we find the NUT-charged solution, compute its curvature, and examine the asymptotic region along the symmetry axis. We conclude the paper with a brief discussion of our results in Sec.~\ref{sc:con}.


\section{Preliminaries}\label{sc:pre}


\subsection{Infinite derivative gravity}
The most general four-dimensional (parity-invariant and torsionless) gravity action that is quadratic in curvature can be written in the form \cite{Biswas:2011ar,Biswas:2016etb,Biswas:2016egy}\footnote{We use the geometric unit system in which ${c=1}$ and ${G=1}$, and mostly positive metric signature ${(-,+,+,+)}$.}
\begin{equation}\label{eq:SQ}
\begin{aligned}
\mathcal{S}_\textrm{QG} &=\frac{1}{16\pi}\int_M\!\!\sqrt{-\mathfrak{g}}\Big[R+\frac12\big(R F_1(\Box)R
\\
&\feq+R_{\mu\nu}F_2(\Box)R^{\mu\nu}+R_{\mu\nu\kappa\lambda}F_3(\Box)R^{\mu\nu\kappa\lambda}\big)\Big]\;,
\end{aligned}
\end{equation}
where the \textit{form-factors} $F_i(\Box)$ are given by the analytic functions of d'Alembertian ${\Box=\nabla_\mu\nabla^\mu}$. In what follows, we focus on the lowest order expansion of this action around Minkowski background ${\eta_{\mu\nu}=\mathrm{diag}(-1,1,1,1)}$ in Cartesian coordinates ${(t,x,y,z)}$,
\begin{equation}\label{eq:metriclinearization}
g_{\mu\nu}=\eta_{\mu\nu}+ h_{\mu\nu}\;, \quad |h_{\mu\nu}| \ll 1\;.
\end{equation}
Note, that we can freely set ${F_3(\Box)=0}$ because all the second order perturbations in $h_{\mu\nu}$ can be absorbed in terms involving $F_1(\Box)$ and $F_2(\Box)$ (see, e.g., \cite{Frolov:2015usa}). Let us further assume that
\begin{equation}
1-F_1(\Box)\Box=1+\frac12 F_2(\Box)\Box = e^{-\Box/M_\textrm{s}^2}\;.
\end{equation}
This choice leads to a particular simple example of the \textit{infinite derivative gravity} \cite{Biswas:2011ar},
\begin{equation}\label{eq:idgaction}
\mathcal{S}_\textrm{IDG}=\frac{1}{16\pi}\int_M\!\!\sqrt{-\mathfrak{g}}\Big[R+G_{\mu\nu}\frac{e^{-\Box/M_\textrm{s}^2}-1}{\Box}R^{\mu\nu}\Big]\;.
\end{equation}
The non-local exponential operator guarantees that this theory is \textit{ghost-free} and has the same number of perturbative degrees of freedom as the general relativity. Indeed, the propagator of the infinite derivative gravity in the Fourier space,\footnote{
Our convention for $n$-dimensional Fourier transform is:
\begin{equation*}
\begin{aligned}
\mathcal{F}[f](k) &=\frac{1}{(2\pi)^{n/2}}\int_{\mathbb{R}^n}\!\!\!d^n x\,f(x)e^{-i k \bdot x}\;,
\\
\mathcal{F}^{-1}[f](x) &=\frac{1}{(2\pi)^{n/2}}\int_{\mathbb{R}^n}\!\!\!d^n \mathrm{k}\,f(k) e^{i k \bdot x}.
\end{aligned}
\end{equation*}
}
\begin{equation}
\Pi_\textrm{IDG}(k)=\frac{\Pi_\textrm{GR}(k)}{e^{k^2/M_\mathrm{s}^2}}\;,
\end{equation}
has the same poles as the propagator of the general relativity ${\Pi_\textrm{GR}(k)}$ since the exponential function is an entire function with no zeros in the complex plane. Therefore, the only propagating degree of freedom is the massless spin-2 graviton corresponding to the pole ${k^2=0}$. The action \eqref{eq:idgaction} reduces to the Einstein--Hilbert action of the general relativity in the \textit{local limit} ${M_{s}\to\infty}$. The equation of motion of the infinite derivative gravity to the first order in metric perturbation ${h_{\mu\nu}}$ is
\begin{equation}\label{eq:fieldequa}
\begin{aligned}
e^{-\Box/M_\textrm{s}^2}[ &\Box h_{\mu\nu}+\eta_{\mu\nu}\pp_\rho\pp_\sigma h^{\rho\sigma}+\pp_\mu\pp_\nu h
\\
&-\pp_\sigma(\pp_\nu h_\mu^\sigma+\pp_\mu h_\nu^\sigma)-\eta_{\mu\nu}\Box h]=-16\pi T_{\mu\nu}\;,
\end{aligned}
\end{equation}
where $T_{\mu\nu}$ is the stress-energy tensor. Imposing the harmonic gauge condition,
\begin{equation}\label{eq:harmgauge}
\pp_\mu h_\nu^\mu=\frac12 \pp_\nu h\;,
\end{equation}
the field equations \eqref{eq:fieldequa} take the form
\begin{equation}\label{eq:linidgharm}
e^{-\Box/M_\textrm{s}^2}\Box h_{\mu\nu}=-16\pi T_{\mu\nu}\;.
\end{equation}
It differs from the linearized general relativity by the additional non-local form-factor ${e^{-\Box/M_\textrm{s}^2}}$, which disappears in the local limit ${M_\textrm{s}\to\infty}$.


\subsection{Taub--NUT spacetime}

Let us consider the metric of the \textit{(massless) Taub--NUT spacetime} \cite{Taub:1950ez,Newman:1963yy} (see also \cite{Griffiths:2009dfa} and references therein),
\begin{equation}\label{eq:nutmetric}
\begin{aligned}
\bs{g} &= -\frac{r^2-N^2}{r^2 +N^2}\big(\bs{\dd}t +2 N(C+\cos\vartheta)\,\bs{\dd}\varphi\big)^2
\\
&\feq+\frac{r^2 +N^2}{r^2-N^2}\,\bs{\dd}r^2 + (r^2 + N^2)(\bs{\dd}\vartheta^2+\sin^2\vartheta\,\bs{\dd} \varphi^2)\;,
\end{aligned}
\end{equation}
where $N$ is the \textit{NUT charge}. The parameter $C$ characterizes the location of the \textit{Misner string} describing a topological defect resembling the spinning semi-infinite cosmic string. The choice ${C=1}$, considered in this paper, corresponds to the Misner string located at the top semi-axis ${\vartheta=0}$. The bottom part of the axis ${\vartheta=\pi}$ is regular for this choice.\footnote{The choice ${C=-1}$ describes a spacetime with the Misner string at the bottom semi-axis, and ${C=0}$ a symmetrical placement of two counter-rotating Misner strings on both semi-axes.} Because of this topological defect, the spacetime cannot be asymptotically flat. The analytically extended geometry has two horizons ${r=\pm |N|}$. Apart from the Misner string the non-linear Taub--NUT geometry has no scalar curvature singularity.

However, we are more interested in the linearized version of the Taub--NUT spacetime. If we rewrite \eqref{eq:nutmetric} with ${C=1}$ in the Cartesian coordinates ${(t,x,y,z)}$,
\begin{equation}
\begin{aligned}
x &=r \sin\vartheta \, \cos\varphi\;,
\\
y &=r \sin\vartheta \, \sin\varphi\;,
\\
z &=r \cos\vartheta\;,
\end{aligned}
\end{equation}
and expand it to the first order in the NUT charge $N$, we obtain \eqref{eq:metriclinearization} with 
\begin{equation}\label{eq:linnut}
h_{tx} = \frac{2 N y}{\rho^2}\bigg(1+\frac{z}{r}\bigg)\;,
\quad
h_{ty} = -\frac{2 N x}{\rho^2}\bigg(1+\frac{z}{r}\bigg)\;,
\end{equation}
where we employ the short notation ${\rho^2=x^2+y^2}$ and ${r^2=x^2+y^2+z^2}$. The linearized metric \eqref{eq:linnut} describes a spacetime with a small NUT charge. It has a scalar curvature singularity at the origin ${r=0}$ and distributional curvature at the symmetry axis for ${z>0}$ due to the presence of the Misner string. Following the Bonnor's interpretation, one can show (see, e.g., \cite{Argurio:2009xr}) that this geometry can be generated by the stress-energy tensor
\begin{equation}\label{eq:nutset}
T_{tx}=-\frac{N}{2}\delta(x)\delta'(y)\theta(z)\;,
\quad
T_{ty}=\frac{N}{2}\delta'(x)\delta(y)\theta(z)\;,
\end{equation}
which corresponds to the semi-infinite spinning cosmic string with no tension. To show this, we consider the Minkowski spacetime in cylindrical coordinates ${(\tau,\rho,\varphi,z)}$,
\begin{equation}
\bs{g}=-\bs\dd \tau^2 + \bs\dd \rho^2+\rho^2 \bs\dd\varphi^2+\bs\dd z^2\;,
\end{equation}
where we assume that the points ${(\tau,\rho,\varphi=0,z)}$ are identified with ${(\tau+8\pi J,\rho,\varphi=2\pi,z)}$. This identification can be reformulated by introducing the smooth temporal coordinate
\begin{equation}
    t=\tau-4J\varphi\;.
\end{equation}
The above construction gives rise to the \textit{spinning cosmic string spacetime} \cite{Deser:1983tn,Mazur:1986gb} of infinite length, angular momentum $J$, and zero tension,
\begin{equation}\label{eq:scsmetric}
\bs{g}=-(\bs\dd t+4J\bs\dd \varphi)^2+\bs\dd \rho^2+\rho^2 \bs\dd\varphi^2+\bs\dd z^2\;.
\end{equation}
This spacetime is locally flat everywhere except for the axis of symmetry. It differs from the Minkowski spacetime is in the presence of the topological defect at the symmetry axis, which changes the global properties of the geometry. (For instance, the spacetime admits closed time-like curves in the region ${\rho<2|J|}$, where the Killing vector ${\bs{\pp}_\varphi}$ is timelike.\footnote{For further details on the spinning cosmic strings and related topological defects, we refer the reader to \cite{Jensen:1992wj,Galtsov:1993ne,Tod:1994,Puntigam:1996vy,Vilenkin:2000jqa}.}) Let us write the metric \eqref{eq:scsmetric} in the Cartesian coordinates,
\begin{equation}
x=\rho\cos\varphi\;,
\quad
y=\rho\sin\varphi\;.
\end{equation}
and expand it to the first order in angular momentum $J$. The resulting linearized metric \eqref{eq:metriclinearization} takes the form
\begin{equation}\label{eq:linscs}
h_{tx}=\frac{4Jy}{\rho^2}\;,
\quad
h_{ty}=-\frac{4Jx}{\rho^2}\;,
\end{equation}
which can be generated by the stress-energy tensor
\begin{equation}\label{eq:scsset}
T_{tx}=-\frac{J}{2}\de(x)\de'(y)\;,
\quad
T_{ty}=\frac{J}{2}\de'(x)\de(y)\;.
\end{equation}
Comparing \eqref{eq:nutset} with \eqref{eq:scsset}, we see that the source for the weakly NUT-charged Taub--NUT spacetime is equivalent to the source of the slowly spinning semi-infinite string with angular momentum ${N}$ localized on the top semi-axis, ${z>0}$. It is also not surprising that \eqref{eq:linscs} can be obtained directly from \eqref{eq:linnut} by taking the asymptotic limit ${z\to\infty}$.

Finally, it is worth noting that the Taub--NUT spacetime and the spinning cosmic string spacetime are also related at the full non-linear level. Particularly, taking the limit ${N\to 0}$, ${NC=2J=\textrm{const.}}$, of \eqref{eq:nutmetric}, we obtain exactly \eqref{eq:scsmetric}. In what follows, we focus solely on the linearized metrics.


\section{NUT-charged source in infinite derivative gravity}\label{sc:nut}

\subsection{Metric}
In order to solve the linearized infinite derivative gravity equations \eqref{eq:linidgharm} we use the following ansatz:
\begin{equation}\label{eq:ansatz}
\begin{aligned}
\bs{g} &={-}\bs{\dd}t^2{+}\bs{\dd}x^2{+}\bs{\dd}y^2{+}\bs{\dd}z^2{+}X\bs{\dd}t\bvee\bs{\dd}x{+}Y\bs{\dd}t\bvee\bs{\dd}y\;,
\end{aligned}
\end{equation}
where ${X=X(x,y,z)}$ and ${Y=Y(x,y,z)}$. Furthermore, the functions $X$ and $Y$ are subject to the harmonic gauge \eqref{eq:harmgauge},
\begin{equation}\label{eq:XYcond}
\pp_x X+\pp_y Y=0\;.
\end{equation}
Using this ansatz, the field equations \eqref{eq:linidgharm} reduce to
\begin{equation}
\begin{aligned}
e^{-\Delta/M_\textrm{s}^2}\Delta X &=8\pi N\de(x)\de'(y)\theta(z)\;,
\\
e^{-\Delta/M_\textrm{s}^2}\Delta Y &=-8\pi N\de'(x)\de(y)\theta(z)\;,
\end{aligned}
\end{equation}
where we denote the three-dimensional Laplacian by ${\Delta\equiv\pp_x^2+\pp_y^2+\pp_z^2}$. We can get rid of the non-local exponential operator by going to the Fourier space ${(k_x,k_y,k_z)}$,
\begin{equation}
\begin{aligned}
e^{(k_x^2+k_y^2+k_z^2)/M_\textrm{s}^2} \mathcal{F}[\Delta X] &= \frac{2\sqrt{2} N k_y}{\sqrt{\pi}}\bigg[\textrm{p.v.}\frac{1}{k_z}+ i \pi  \delta (k_z)\bigg]\;,
\\
e^{(k_x^2+k_y^2+k_z^2)/M_\textrm{s}^2}\mathcal{F}[\Delta Y] &=-\frac{2 \sqrt{2} N k_x}{\sqrt{\pi}}\bigg[\textrm{p.v.}\frac{1}{k_z}+ i \pi\delta (k_z)\bigg]\;\!.
\end{aligned}
\end{equation}
If we now divide both sides of these two equations by $e^{(k_x^2+k_y^2+k_z^2)/M_\textrm{s}^2}$ and take the inverse Fourier transform, we arrive at the Poisson equations
\begin{equation}
\begin{aligned}
\Delta X &=-\frac{N M_\textrm{s}^4 y}{2} \big(1+\erf(M_\textrm{s} z/2)\big) e^{-M_\textrm{s}^2 \rho^2/4}\;,
\\
\Delta Y &=\frac{N M_\textrm{s}^4 x}{2} \big(1+\erf(M_\textrm{s} z/2)\big) e^{-M_\textrm{s}^2 \rho^2/4}\;.
\end{aligned}
\end{equation}

Considering the axial symmetry of the problem, we can assume that the solution takes the form
\begin{equation}\label{eq:XYusingV}
\begin{aligned}
X(x,y,z) &=\frac{y}{\rho^2}V(\rho^2/4,z)\;,
\\
Y(x,y,z) &=-\frac{x}{\rho^2}V(\rho^2/4,z)\;,
\end{aligned}
\end{equation}
which automatically satisfy the condition \eqref{eq:XYcond}. The particular choice of the radial dependence allows us to rewrite the two 3-dimensional Poisson equations as one 2-dimensional partial differential equation (of Keldysh type \cite{Otway2012}) for function ${V=V(v,z)}$ of variables ${v=\rho^2/4}$ and $z$,
\begin{equation}
(v\pp_v^2 + \pp_z^2) V = -2N M_\textrm{s}^4\big(1+\erf(M_\textrm{s} z /2)\big)v e^{-M_\textrm{s}^2 v}\;.
\end{equation}
We can reduce the degree of the derivative in $v$ by taking the Laplace transform in variable $v$,\footnote{We use the following convention for the Laplace transform:
\begin{equation*}
\begin{aligned}
\mathcal{L}[f](s) &=\int_0^\infty\!\!\!d v\,f(v)e^{-sv}\;,
\\
\mathcal{L}^{-1}[f](v) &=\frac{1}{2\pi i}\int_{\gamma-i\infty}^{\gamma+i\infty}\!\!\!d s\,f(s)e^{sv}\;.
\end{aligned}
\end{equation*}
where ${\gamma>0}$ is an arbitrary constant.
} 
\begin{equation}\label{eq:Vlaplequat}
(s^2\pp_s+2s-\pp_z^2)\mathcal{L}[V](s;z)
=\frac{2 M_\mathrm{s}^4 N}{(s{+}M_\mathrm{s}^2)^2}\big(1{+}\erf(M_\mathrm{s}z/2)\big)\;,
\end{equation}
where we set the boundary condition at the symmetry axis ${V(0,z)=0}$, because we are interested in solutions for which $X$ and $Y$ are finite. Employing the substitution
\begin{equation}
U(t,z)=\frac{1}{t^2}\mathcal{L}[V]\Big({-}\frac{1}{t};z\Big)\;,
\end{equation}
it is simple to show that \eqref{eq:Vlaplequat} can be cast in the form of the heat equation
\begin{equation}\label{eq:inhomheat}
(\pp_t - \pp_z^2) U=\frac{2 M_\mathrm{s}^4 N}{(M_\mathrm{s}^2 t-1)^2} \big(1+\erf(M_\mathrm{s}z/2)\big)\equiv \Phi\;.
\end{equation}

We solve this equation by the method of \textit{Green's function} ${G(t,z;\tilde{t},\tilde{z})}$, which satisfies
\begin{equation}
(\pp_t - \pp_z^2) G(t,z;\tilde{t},\tilde{z})=\de(t-\tilde{t})\de(z-\tilde{z})\;.
\end{equation}
The function satisfying this equation is given by the expression
\begin{equation}
G(t,z;\tilde{t},\tilde{z})=\frac{e^{-(z-\tilde{z})^2/4(t-\tilde{t})}}{\sqrt{4\pi(t-\tilde{t})}}\theta(t-\tilde{t})\;,
\end{equation}
which is commonly referred to as the \textit{heat kernel}. The solution of the inhomogeneous equation \eqref{eq:inhomheat} is then given by the right-hand side $\Phi$ with the heat kernel,
\begin{equation}
U(t,z)=\int_{\mathbb{R}^2}\!\!\!d\tilde{t}d\tilde{z}\,G(t,z;\tilde{t},\tilde{z})\Phi(\tilde{t},\tilde{z})\;,
\end{equation}
or explicitly,
\begin{equation}
U(t,z) = \frac{4 M_\mathrm{s}^3 N}{\sqrt{\pi}}\int_0^\infty\!\!\! d p\!\int_\mathbb{R}\!\!\! d q \frac{\big(1{+}\erf(q)\big)e^{-(q/M_\mathrm{s}p - z/2p)^2}}{(1+M_\mathrm{s}^2(p^2-t))^2} \;,
\end{equation}
where we used substitutions ${p=\sqrt{t-\tilde{t}}}$ and ${q=M_\mathrm{s}\tilde{z}/2}$. The inner integral can be found with the help of the identity \cite{Ng:1969}
\begin{equation}
\int_\mathbb{R} \!\!dq\, \erf(q)\,e^{-(a q+b)^2} = -\frac{\sqrt{\pi}}{a}\erf\Big(\frac{b}{\sqrt{1+a^2}}\Big)
\end{equation}
and the standard formula for the integral of the Gaussian function, ${\int_\mathbb{R} \!dq\, e^{-(a q+b)^2} {=} \sqrt{\pi}/a}$. We arrive at the integral
\begin{equation}
U(t,z)=4 M_\mathrm{s}^4 N\int_0^\infty\!\!\! d p \,\frac{p\Big[1+\erf\Big(\frac{M_\mathrm{s}z}{2\sqrt{1+M_\mathrm{s}^2 p^2}}\Big)\Big]}{\big(1+M_\mathrm{s}^2(p^2-t)\big)^2}\;.
\end{equation}

It turns out that it is much easier to first calculate the simple inverse Laplace transform of ${U(-1/s,z)/s^2}$ and then deal with the integration in $p$. Interchanging the integration in $p$ with the integration in the Laplace variable $s$, we can write $V(v,z)$ as follows:
\begin{equation}
V(v,z) = 4 M_\mathrm{s}^4 N v \!\!\int_0^\infty\!\!\!\!\! d p  \,\frac{p\Big[1{+}\erf\!\Big(\frac{M_\mathrm{s}z}{2\sqrt{1{+}M_\mathrm{s}^2 p^2}}\Big)\Big]e^{-\frac{M_\mathrm{s}^2 v}{1{+}M_\mathrm{s}^2p^2}}}{\big(1+M_\mathrm{s}^2p^2\big)^2}\;\!.
\end{equation}
This integral may look intimidating at first, but it can be found in a few simple steps.\footnote{The primitive function is obtained using the integration by parts where we differentiate the expression in the square bracket. The two integrals that need to be evaluated are then of the form ${\int\! d p\, \exp(-f(p))f'(p)}$ and ${\int\! d p\, \exp(-g(p)^2)g'(p)}$ that is suitable for the integration by substitution.} After returning to the standard radial variables ${\rho=2\sqrt{v}}$ and ${r}$, we obtain a very compact result,
\begin{equation}\label{eq:Vsolution}
V=2 N \Big[1+\frac{z}{r}\erf(M_\textrm{s} r/2)- \big(1{+}\erf(M_\mathrm{s}z/2)\big)e^{-M_\textrm{s}^2 \rho^2/4}\Big]\;.
\end{equation}

\begin{figure*}
  \centering
      \includegraphics[width=\columnwidth/2]{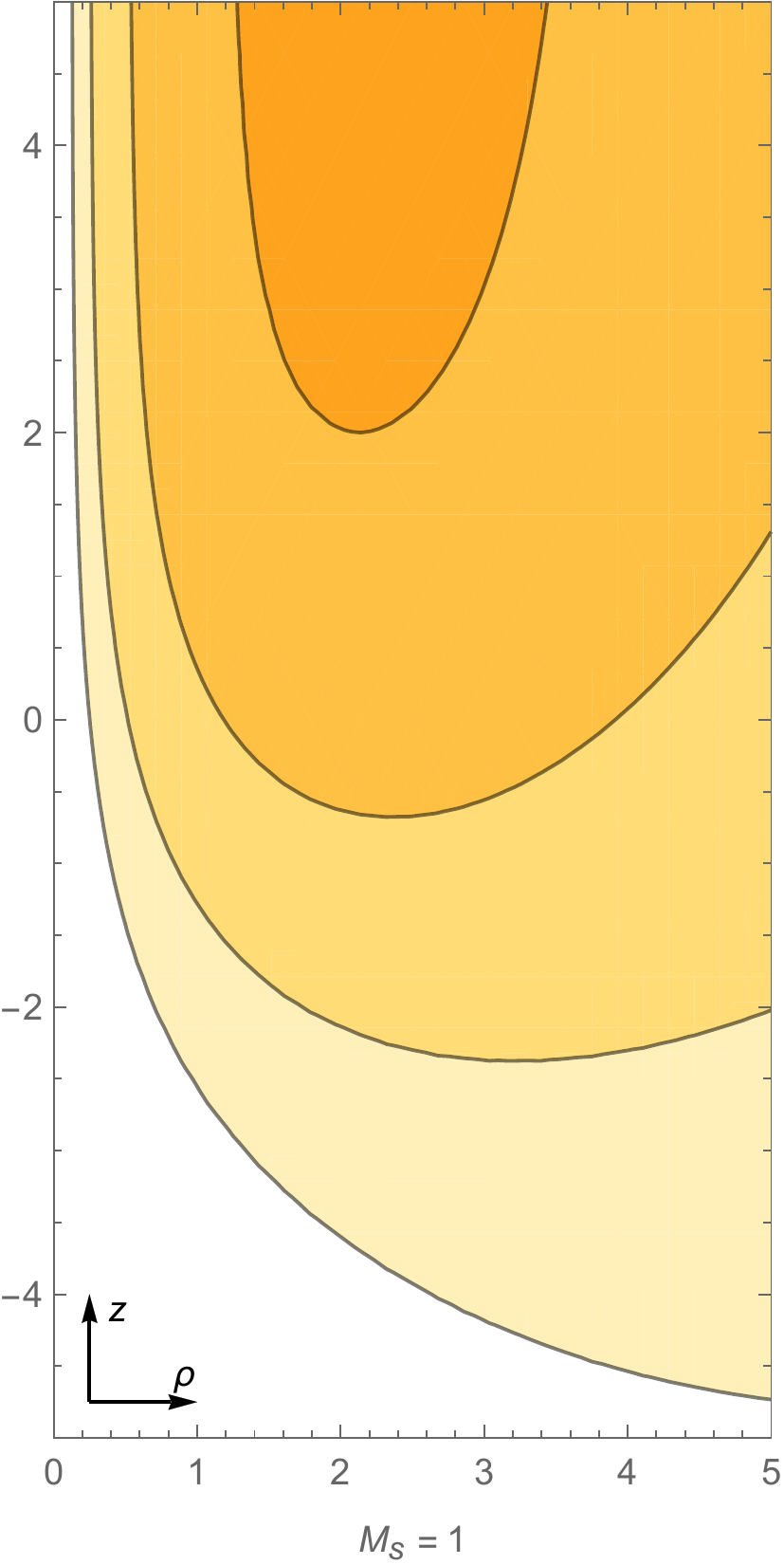}
      \includegraphics[width=\columnwidth/2]{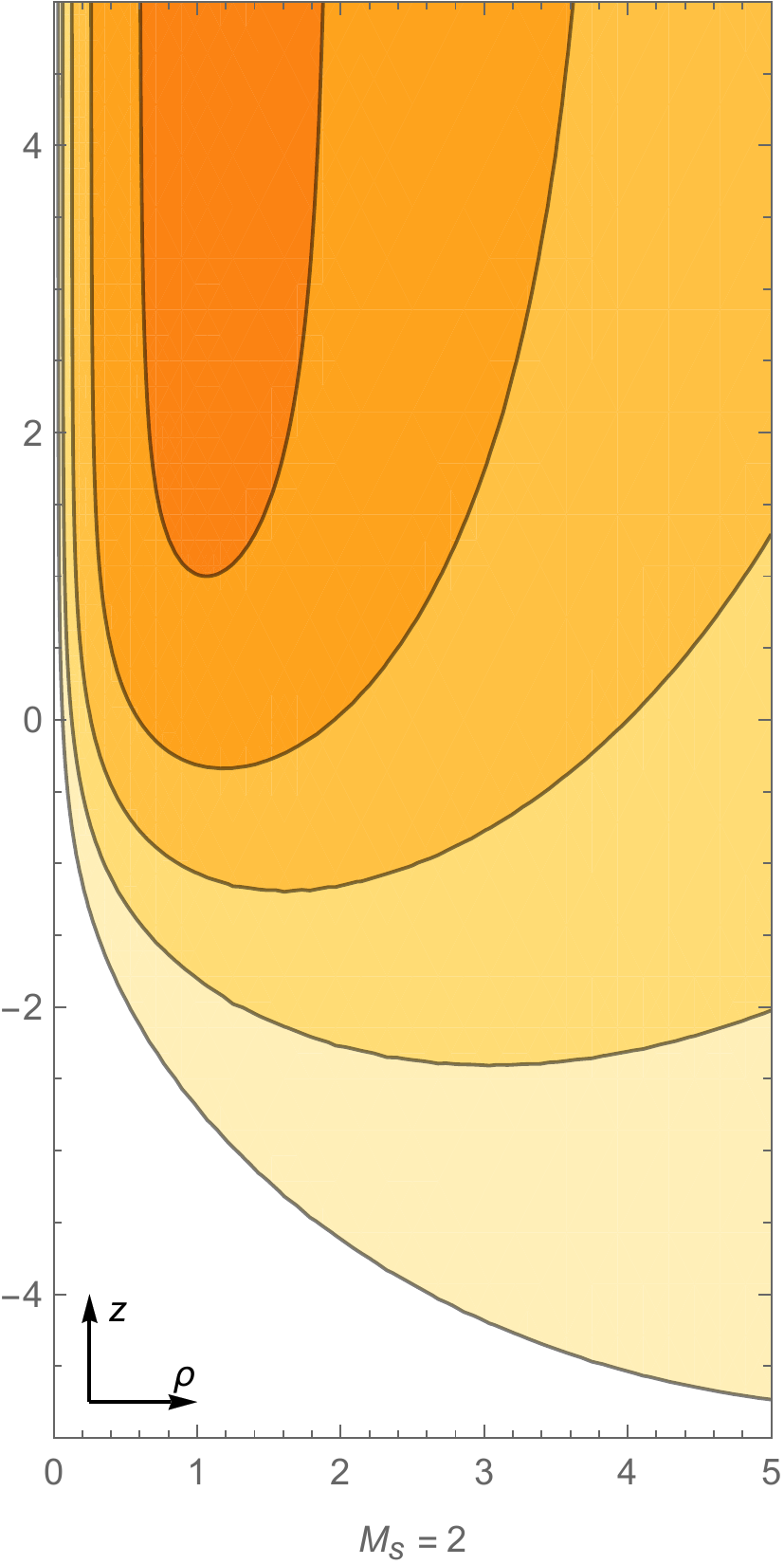}
      \includegraphics[width=\columnwidth/2]{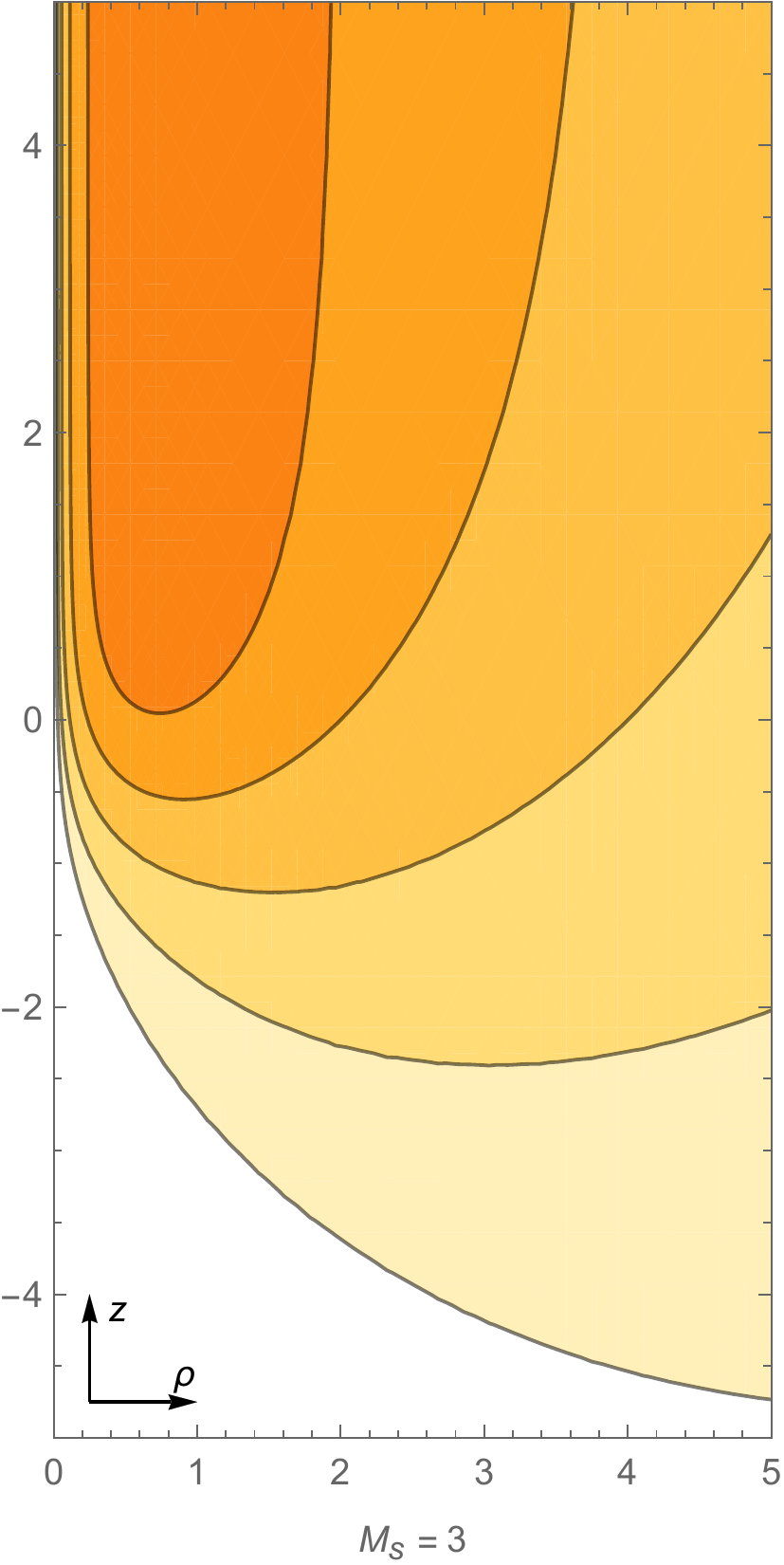}
      \includegraphics[width=\columnwidth/2]{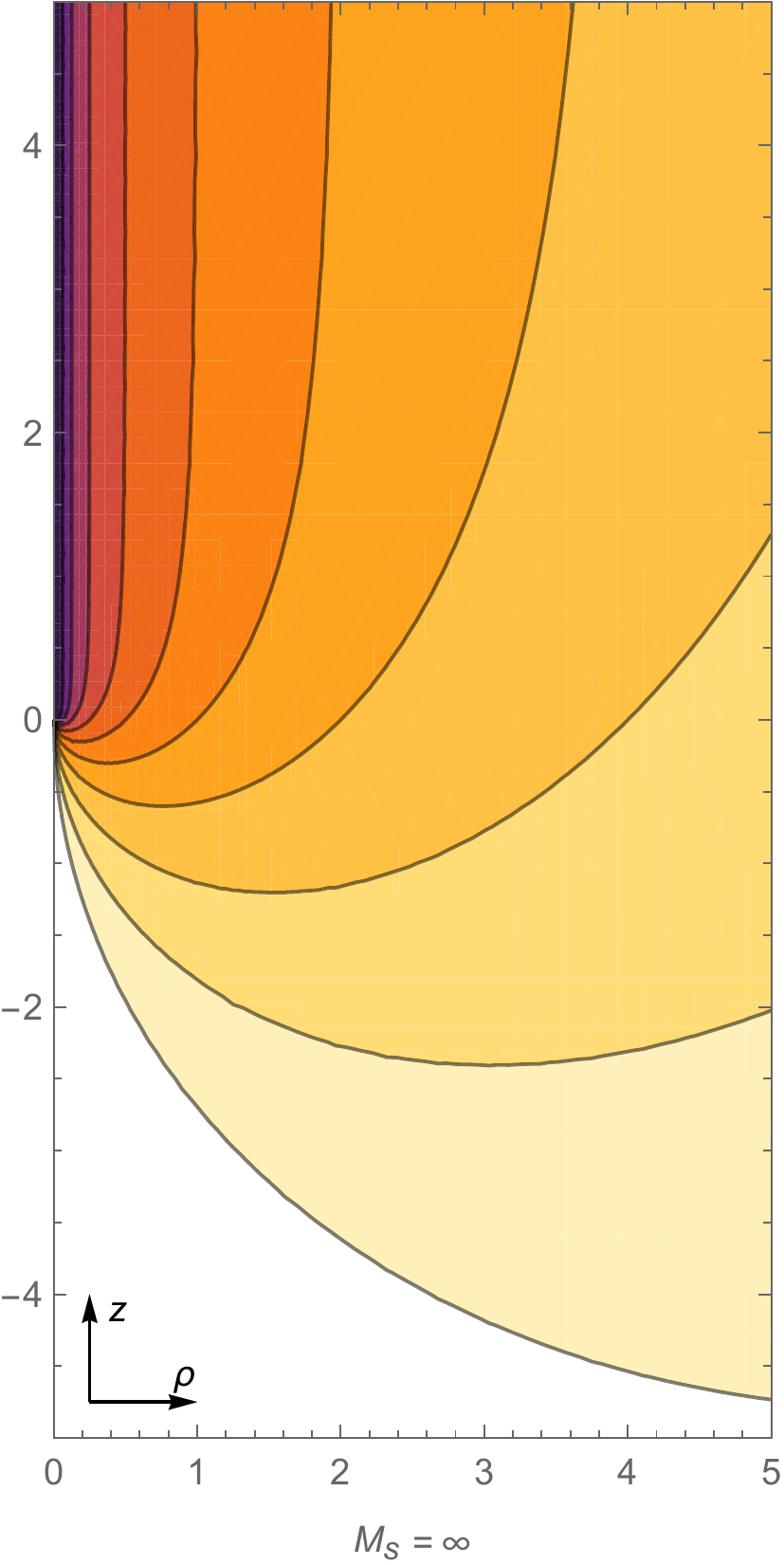}
      \includegraphics[width=\columnwidth]{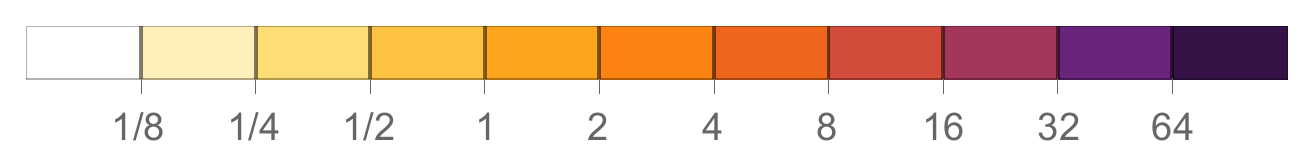}
  \caption{The metric function of the NUT-charged spacetime in the infinite derivative gravity. We plot the contours of ${V/\rho N}$ for different values of the non-local scale $M_\textrm{s}$. The function is finite everywhere for finite values of $M_\textrm{s}$, but diverges toward the top semi-axis (the Misner string) in the local limit, ${M_\textrm{s}=\infty}$.}\label{fig:vrhonut}
\end{figure*}

Far from the axis the geometry approaches the linearized Taub--NUT solution of the general relativity \eqref{eq:linnut} because
\begin{equation}
V \approx 2 N \Big(1+\frac{z}{r}\Big)\equiv V_\textrm{loc}\;, \quad \rho\gg1/M_\textrm{s}\;.
\end{equation}
The Taub--NUT spacetime is also reproduced in the local limit ${M_\textrm{s}\to\infty}$,
\begin{equation}
\lim_{M_\textrm{s}\to\infty}V = V_\textrm{loc}\;, \quad \rho>0\;.
\end{equation}
The functions ${X}$ and ${Y}$,
\begin{equation}
X=\frac{y}{\rho^2} V\;, 
\quad
Y=-\frac{x}{\rho^2} V\;,
\end{equation}
are finite everywhere unlike the corresponding ${X_\textrm{loc}}$ and ${Y_\textrm{loc}}$ which diverge at ${\rho=0}$, for ${z\ge 0}$ since 
\begin{equation}
\lim_{\rho\to 0} V/\rho = 0\;, \quad \lim_{\rho\to 0} V_\textrm{loc}/\rho = \infty\;.
\end{equation}
(The expressions ${x/\rho=\cos{\varphi}}$ and ${y/\rho=\sin{\varphi}}$ are finite and non-zero for general ${\varphi}$.) This already hints that there is no Misner string present in the non-local case. The contour plots of the function ${V/\rho N}$ are shown in Fig.~\ref{fig:vrhonut}.

\begin{figure*}
  \centering
      \includegraphics[width=\columnwidth/2]{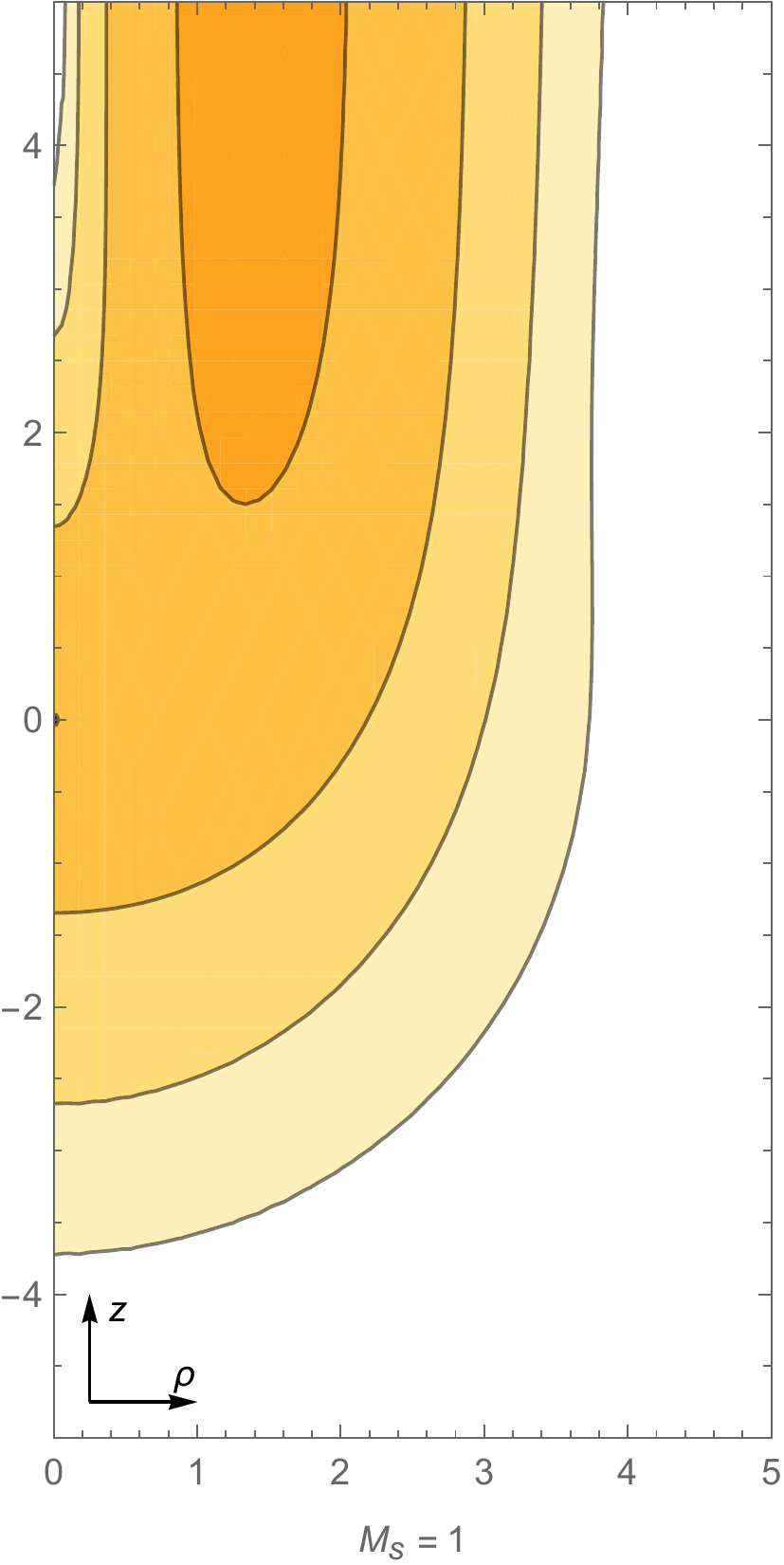}
      \includegraphics[width=\columnwidth/2]{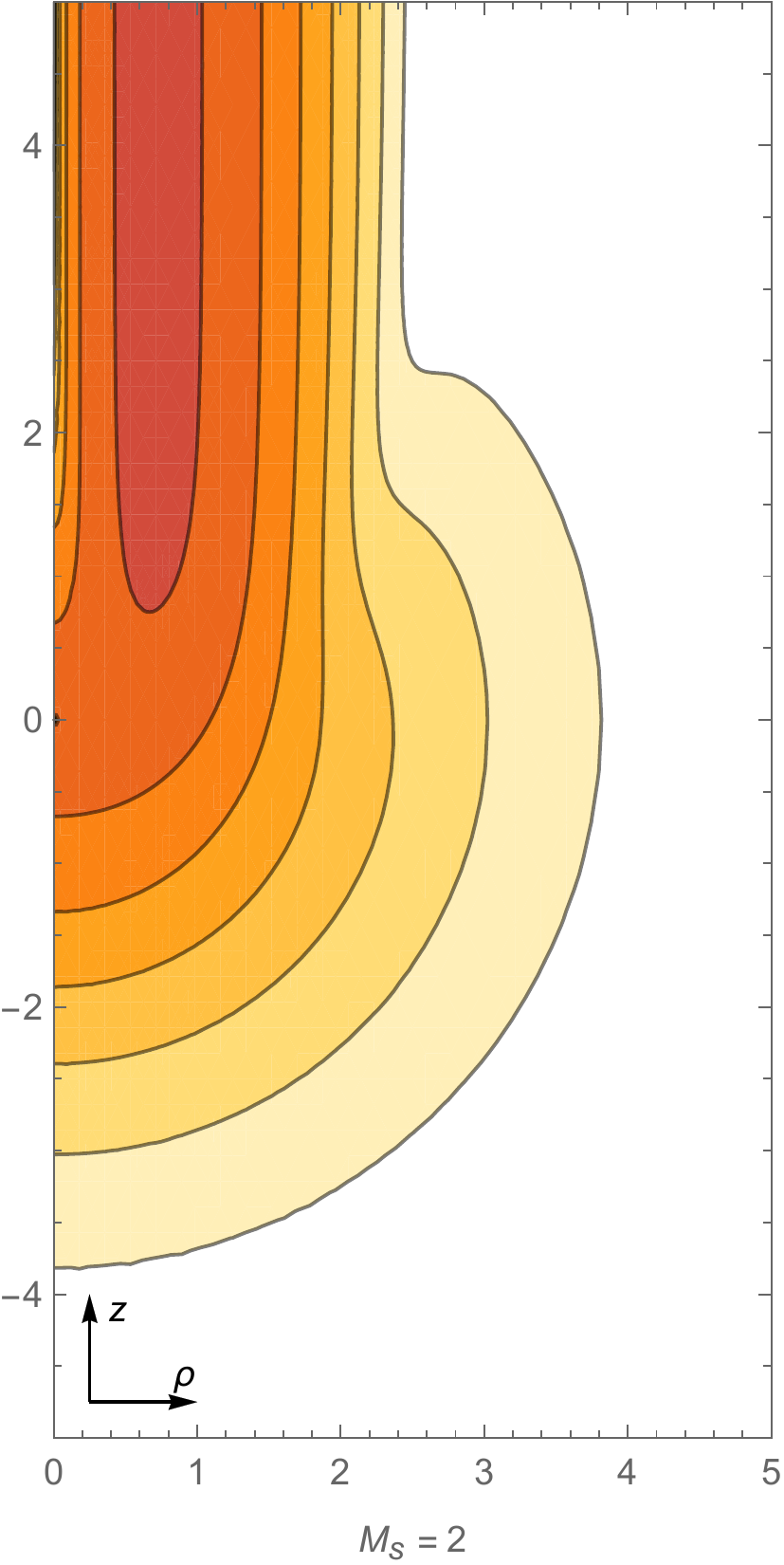}
      \includegraphics[width=\columnwidth/2]{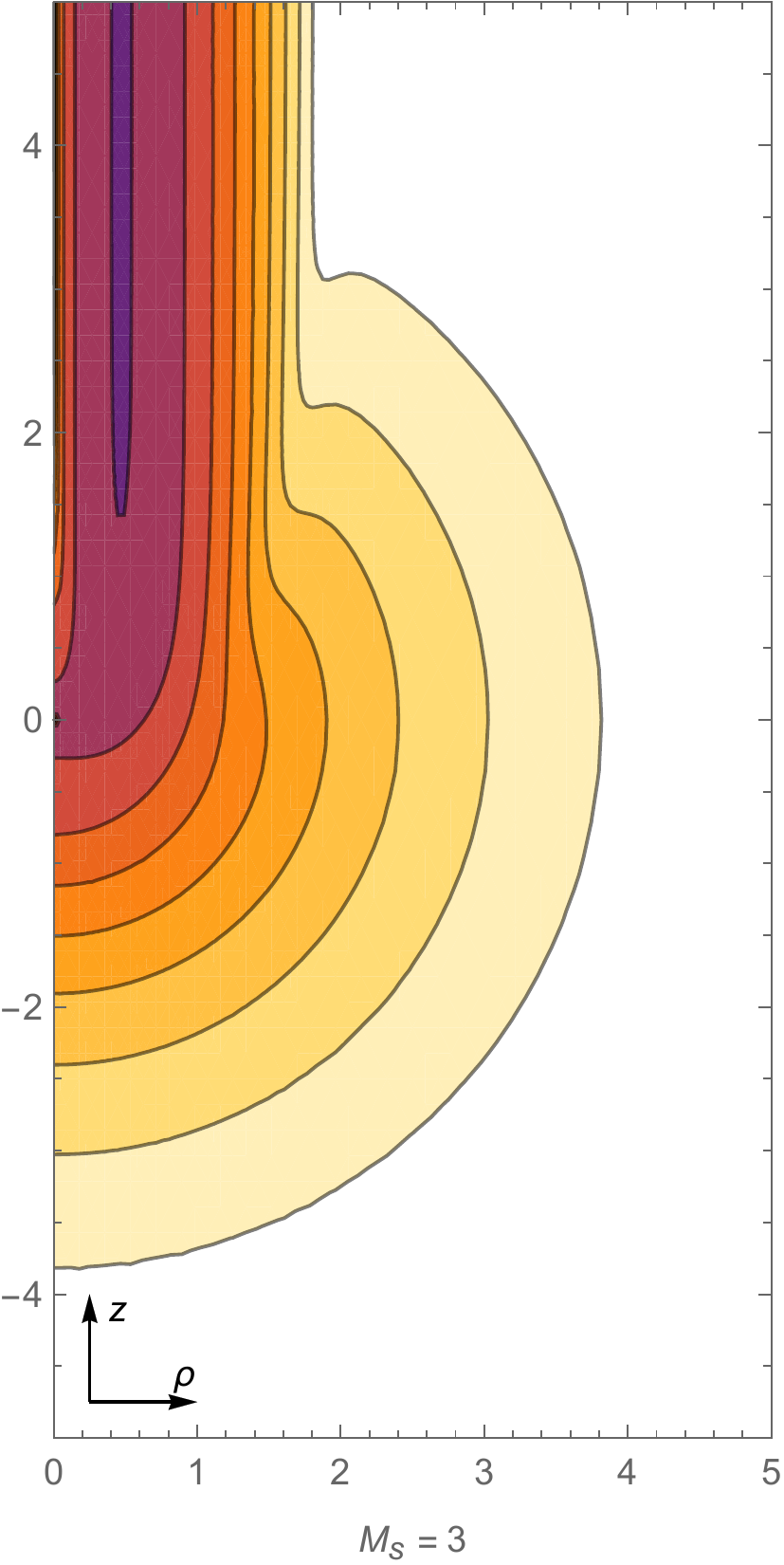}
      \includegraphics[width=\columnwidth/2]{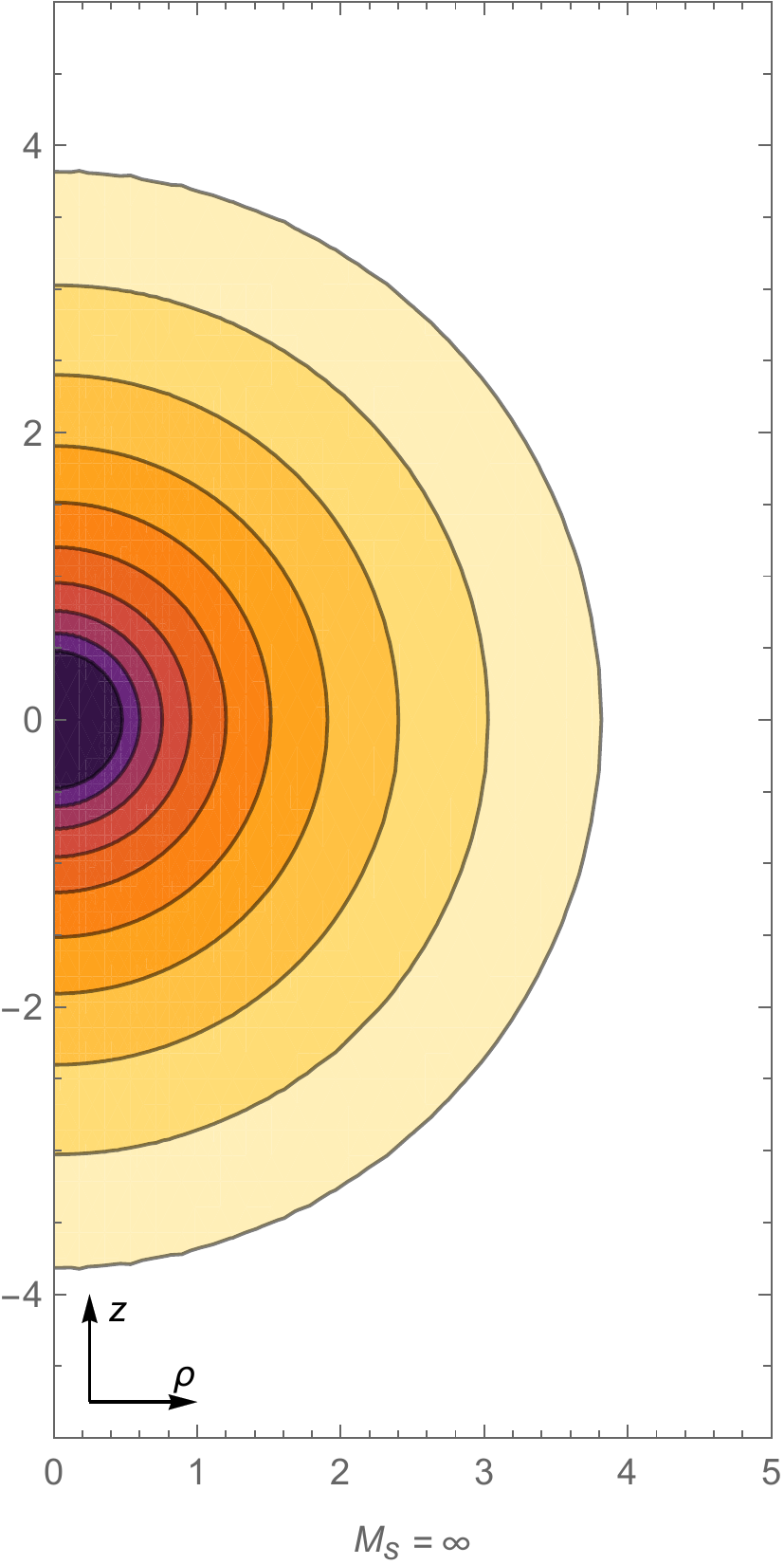}
      \includegraphics[width=\columnwidth]{leg}
  \caption{The Kretschmann scalar of the NUT-charged spacetime in the infinite derivative gravity. We plot the contours of $\sqrt{-K}/|N|$ for different values of the non-local scale $M_\textrm{s}$. The Kretschmann scalar is finite everywhere for finite $M_\textrm{s}$. With the increasing value of $M_\textrm{s}$, the curvature accumulates along the top semi-axis and gives rise to distributional curvature (the Misner string) in the local limit, ${M_\textrm{s}=\infty}$, for which the Kretschmann scalar also diverges at the origin.}\label{fig:kretnut}
\end{figure*}


\subsection{Curvature}
To the first order in metric perturbation ${h_{\mu\nu}}$, the Riemann tensor, the Ricci tensor, and the Ricci scalar in the harmonic gauge \eqref{eq:harmgauge} read
\begin{equation}
\begin{gathered}
R_{\mu\nu\rho\sigma} =\pp_\nu\pp_{[\rho} h_{\sigma]\mu}-\pp_\mu\pp_{[\rho}h_{\sigma]\nu}\;,
\\\
R_{\mu\nu} =-\frac12\Box h_{\mu\nu}\;,
\quad
R =-\frac12 \Box h \;.
\end{gathered}
\end{equation}
Using the metric ansatz \eqref{eq:ansatz} together with \eqref{eq:XYcond}, we find the non-trivial component expressed in terms of metric functions $X$ and $Y$:
\begin{equation}
\begin{gathered}
\begin{aligned}
R_{txxy}&=\frac12\pp^2 Y\;,
&\quad
R_{tyyx}&=\frac12\pp^2 X\;,
\\
R_{txxz}&=-\frac12\pp_x\pp_z X\;,
&\quad
R_{tyyz}&=-\frac12\pp_y\pp_z Y\;,
\\
R_{txyz}&=-\frac12\pp_x\pp_z Y\;,
&\quad
R_{tyxz}&=-\frac12\pp_y\pp_z X\;,
\\
R_{tzxz}&=-\frac12\pp_z^2 X\;,
&\quad
R_{tzyz}&=-\frac12\pp_z^2 Y\;,
\end{aligned}
\\
R_{tzxy}=\frac12(-\pp_y\pp_z X+\pp_x\pp_z Y)\;,
\\
R_{tx}=-\frac12\Delta X\;,
\quad
R_{ty}=-\frac12\Delta Y\;,
\quad
R=0\;,
\end{gathered}
\end{equation}
where we introduced the two-dimensional Laplacian operator ${\pp^2\equiv\pp_x^2+\pp_y^2}$. 

Note that all the components of the Riemann tensor (including all combinations of covariant and contravariant indices) are finite everywhere. This already implies that there is no curvature singularity or topological defect such as the Misner string in the spacetime. Nevertheless, to get simple invariant information about the curvature, we also compute the Kretschmann scalar ${K=R_{\mu\nu\rho\sigma}R^{\mu\nu\rho\sigma}}$ to the lowest order in metric perturbations. We express it in terms of the function $V$ by means of \eqref{eq:XYusingV},
\begin{equation}\label{eq:kretschXYV}
\begin{aligned}
K &=-2\Big[(\pp^2 X)^2+(\pp^2 Y)^2 +(\pp_z^2 X)^2+(\pp_z^2 Y)^2
\\
&\feq+(\pp_x\pp_z X)^2+(\pp_y\pp_z Y)^2+(\pp_y\pp_z X)^2+(\pp_x\pp_z Y)^2
\\
&\feq + (\pp_y\pp_z X+\pp_x\pp_z Y)^2\Big]
\\
&=-\frac{1}{4v^2}\big[2 v^3 (\pp_v^2 V)^2 + 3 v^2 (\pp_v\pp_z V)^2 + 2 v (\pp_z^2 V)^2
\\
&\feq +(v\pp_v\pp_z V-\pp_z V)^2\big]\;,
\end{aligned}
\end{equation}
where we can observe that ${K\le 0}$. Inserting \eqref{eq:Vsolution} in \eqref{eq:kretschXYV}, we obtain
\begin{equation}
\begin{aligned}
K&=-\frac{2N^2}{\pi\, r^6 z^2}\Big[M_\textrm{s}^6 \rho ^2 r^4 \big(r^2 \beta (\tfrac{M_\textrm{s}z}{2})^2{-}2 \rho ^2 \beta (\tfrac{M_\textrm{s}z}{2}){+}r^2\big)
\\
&\feq+12 z^2 \alpha (\tfrac{M_\textrm{s}r}{2})\big(2\alpha (\tfrac{M_\textrm{s}r}{2}){-} M_\textrm{s}^3\rho ^2 r \beta (\tfrac{M_\textrm{s}z}{2})\big)\Big] e^{-M_\textrm{s}^2 r^2/2}\;,
\end{aligned}
\end{equation}
where $\alpha$ and $\beta$ are functions defined by
\begin{equation}
\begin{aligned}
\alpha(w) &=\sqrt{\pi} \erf(w)e^{w^2}-2 w\;,
\\
\beta(w) &=\sqrt{\pi} \big(1+\erf(w)\big)w e^{w^2}+1\;.
\end{aligned}
\end{equation}
Examining the limits ${r\to0}$ and ${z\to0}$, one can verify that the Kretschmann scalar is also finite everywhere.

It is interesting to investigate the local limit of the Kretschmann scalar. Outside the axis of symmetry we see that it reduces to the Kretschmann scalar of the Taub--NUT spacetime,
\begin{equation}
\lim_{M_\textrm{s}\to\infty} K = K_\textrm{loc}=-\frac{48 N^2}{r^3}\;, \quad \rho>0\;,
\end{equation}
which diverges toward the origin ${r=0}$. Note that this is a feature of the linearized Taub--NUT only.\footnote{The Kretschmann scalar of the full Taub--NUT metric \eqref{eq:nutmetric} is \begin{equation*}
    K_\textrm{loc}=\frac{48 N^2(N^6{-}15 N^4 r^2{+}15 N^2 r^4{-}r^6)}{(N^2{+}r^2)^6}=-\frac{48 N^2}{r^3}+\mathcal{O}(N^4)\;.
\end{equation*}}
The Kretschmann scalar is depicted in the contour plots in Fig.~\ref{fig:kretnut}. Inspecting the graphs for increasing values of the non-local scale ${M_\textrm{s}}$, we see that the curvature really accumulates around the top axis. This can be viewed as a process in which the Misner string corresponding to distributional curvature arises in the local limit.


\subsection{Asymptotic regions: ${z\to\pm\infty}$}

\begin{figure}
  \centering
      \includegraphics[width=\columnwidth]{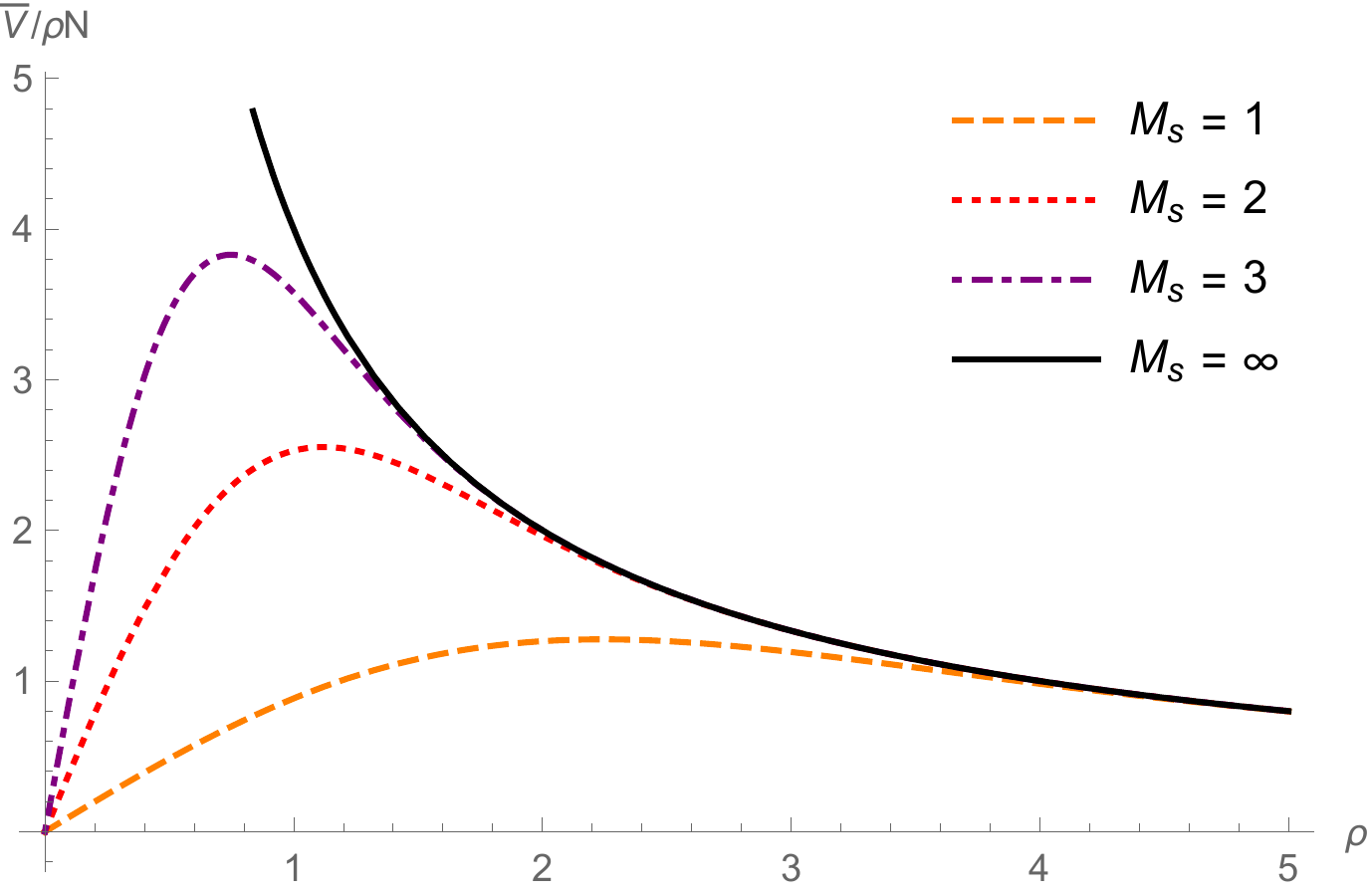}
  \caption{The metric function of the spinning cosmic string spacetime in the infinite derivative gravity. We plot the function ${\overline{V}/\rho N}$ for different values of non-local scale $M_\textrm{s}$. The function is finite everywhere for finite values of ${M_\textrm{s}}$, but diverges toward ${\rho=0}$ in the local limit, ${M_\textrm{s}=\infty}$.}\label{fig:vrhoscs}
\end{figure}

Let us explore asymptotic limits of the spacetime for large positive and negatives values of coordinate $z$, which correspond to the far region around the top and the bottom semi-axes, respectively. In these asymptotic regions, the function $V$ becomes independent of $z$, and it can be simply approximated by its limits,
\begin{equation}\label{eq:scsV}
\lim_{z\to\pm\infty}V=
\begin{cases}
4 N \big(1-e^{-M_\textrm{s}^2 \rho^2/4}\big)\equiv\overline{V}\;,
\\
0\;.
\end{cases}
\end{equation}
The geometry approaches Minkowski spacetime in the bottom part and the spacetime of the spinning cosmic string of an infinite length in the top section. The latter is generated by the infinite linear source \eqref{eq:scsset} with angular momentum ${J=N}$, which was recently found in \cite{Boos:2020kgj}.\footnote{Note that we actually obtained this solution independently of \cite{Boos:2020kgj} as a particular limiting case of our NUT-charged solution before the paper appeared.} The asymptotic geometry given by ${\overline{V}}$ approaches the solution of the local theory \eqref{eq:linscs} with ${J=N}$ far from the axis because
\begin{equation}
    \overline{V}\approx 4N\equiv\overline{V}_\textrm{loc}\;,\quad \rho\gg1/M_\textrm{s}\;.
\end{equation}
Naturally, $\overline{V}$ reduces to this solution in the local limit ${M_\textrm{s}\to\infty}$ as well,
\begin{equation}
\lim_{M_\textrm{s}\to\infty}\overline{V} = \overline{V}_\textrm{loc}\;, \quad \rho>0\;.
\end{equation}
As before, the functions ${\bar{X}}$ and ${\bar{Y}}$ are finite everywhere in contrast to ${\bar{X}_\textrm{loc}}$ and ${\bar{Y}_\textrm{loc}}$. The function ${\overline{V}/\rho N}$ is visualized in Fig.~\ref{fig:vrhoscs}.

\begin{figure}
  \centering
      \includegraphics[width=\columnwidth]{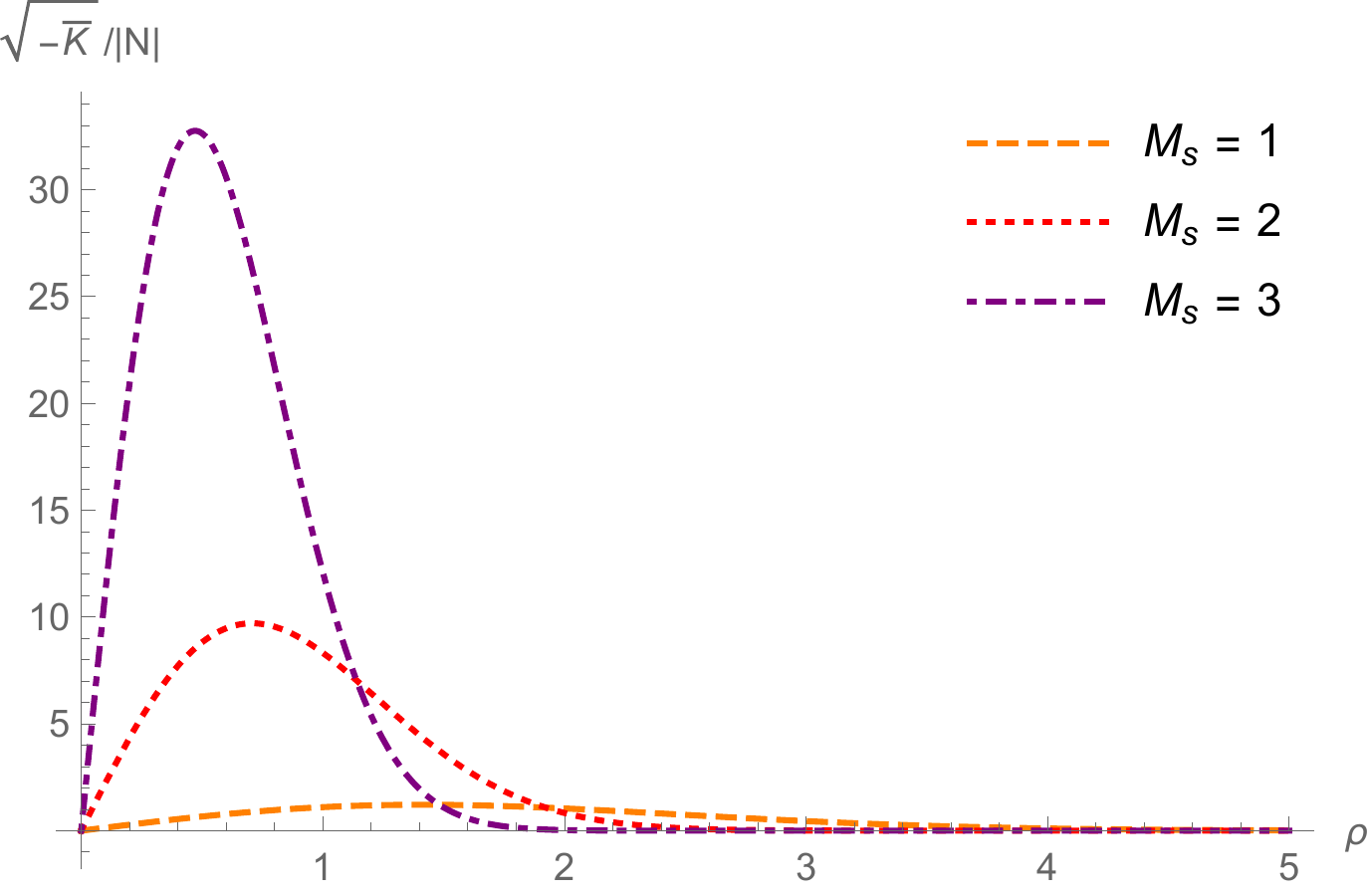}
  \caption{The Kretschmann scalar of the spinning cosmic string spacetime in the infinite derivative gravity. We plot $\sqrt{-\overline{K}}/|N|$ for different values of non-local scale $M_\textrm{s}$. The Kretschman scalar is finite everywhere for finite ${M_\textrm{s}}$. With the increasing value of ${M_\textrm{s}}$, the curvature accumulates at ${\rho=0}$ as described in \eqref{eq:lockretapp}.}\label{fig:kretscs}
\end{figure}

The Kretschmann scalar can be obtained by ignoring the terms with $\pp_z$ in \eqref{eq:kretschXYV},
\begin{equation}\label{eq:scskret}
\begin{aligned}
\overline{K} &=-2\big[(\pp^2 \overline{X})^2+(\pp^2 \overline{Y})^2\big]=-\frac{v}{2}(\pp_v^2 \overline{V})^2
\\
&=-2 M_\textrm{s}^8 N^2 \rho^2 e^{-M_\textrm{s}^2 \rho^2/2}\;.
\end{aligned}
\end{equation}
As expected, it vanishes everywhere outside the axis in the local limit
\begin{equation}
\lim_{M_\textrm{s}\to\infty}\overline{K}= \overline{K}_\textrm{loc} = 0 \;, \quad \rho>0\;,
\end{equation}
because the spacetime of the spinning cosmic string is locally flat for ${\rho>0}$. This analysis again confirms that the curvature really accumulates along the axis in the local limit as one can also see from the graphs of the Kretschmann scalar in Fig.~\ref{fig:kretscs}.

Finally, it is worth to mention that if we approximate the Dirac delta function by the Gaussian function ${\de(x)\approx e^{-x^2/\epsilon^2}/\sqrt{\pi}\epsilon}$, ${0<\epsilon\ll 1}$, then the field equations imply
\begin{equation}\label{eq:lockretapp}
\begin{aligned}
\sqrt{-\overline{K}_\textrm{loc}} &=\sqrt{2\big[(\pp^2 \overline{X}_\textrm{loc})^2{+}(\pp^2 \overline{Y}_\textrm{loc})^2\big]}
\\
&\approx 2^{9/2}|N|\epsilon^{-4}\rho e^{-\rho^2/\epsilon^2}
\\
&\approx -8\sqrt{2} \pi |N|\,\pp_\rho\big(\delta(x)\delta(y)\big)\;, \quad 0<\epsilon\ll 1\;.
\end{aligned}
\end{equation}
Comparing \eqref{eq:lockretapp} with \eqref{eq:scskret}, we can see that ${\epsilon=2/M_\textrm{s}}$, for ${M_\textrm{s}\gg 1}$. This very sloppy calculation indicates how distributional curvature arises in the local limit.


\section{Conclusions}\label{sc:con}

The Taub--NUT spacetime has recently regained significant attention as several presumed unphysical properties have been disproved. In this paper, we studied NUT charge in the context of linearized infinite derivative gravity. We found an analytic solution which is regular everywhere due to the presence of the non-local form-factor, which effectively smears the distributional source. Our analysis of the Riemann tensor and Kretschmann scalar shows the absence of curvature singularities as well as topological defects such as the Misner string. The obtained geometry reduces to the linearized Taub--NUT spacetime far from the source and in the local limit. We also investigated the asymptotic limit along the axis, which gives rise to the non-local version of the spinning infinite string spacetime. Our results extend the set of papers on the linearized solutions by the NUT-charged solution.

A natural next step is to investigate further properties of the NUT charge in infinite derivative gravity. An example is the presence of the closed timelike curves and the existence of the horizons. Unfortunately, these problems are challenging since a full non-linear approach is required. Our knowledge of the exact solutions in this theory is still very sparse, but there are already many hints that might help us make progress in this direction. 

Nevertheless, even at the level of linearized theory, there are still many interesting solutions of the general relativity that might have their counterparts in the infinite derivative gravity. As mentioned in the introduction, the linearized solutions in the non-local theories may actually play more important role than in the local theories.

Our discussion of the presence of singularities is based on the linearized theory. However, it is expected that even the full non-linear theory does not admit singular solutions. The field equations of the full theory involve non-local form-factors acting on the curvature tensors. These operators typically smear the distribution curvature and produce smooth functions. It is anticipated that this mechanism could prevent vacuum solutions (with distributional stress-energy tensor) of local theories with polynomial form-factors to be vacuum solutions of non-local theories. Unfortunately, satisfactory proofs of such statements are still lacking even for very simple examples of singular spacetimes. Nevertheless, the absence of singularities is already partially hinted by the existence of bouncing cosmological solutions \cite{Biswas:2005qr} and approximate spherically symmetric solutions that are valid in the region where non-locality is important \cite{Buoninfante:2018rlq}.


\section*{Acknowledgements}
We would like to thank Pavel Krtou\v{s} for useful discussions about the meaning of the NUT charge. We are also thankful to Shubham Maheshwari for drawing our attention to a recent paper. 

I.K. and A.M. were supported by Netherlands Organization for Scientific Research (NWO) Grant No. 680-91-119.



%

\end{document}